\documentclass[aps,pra,reprint,showpacs,amssymb,amsmath,superscriptaddress,longbibliography,floatfix]{revtex4-1}

\usepackage[english]{babel}
\usepackage{graphicx}
\usepackage{times}
\usepackage{color}
\usepackage{siunitx}
\usepackage{hyperref} 
\usepackage{braket}

\newcommand{\PSDo}{$\mathrm{PSD}_o$}

\DeclareSIUnit\gauss{G}
\DeclareSIUnit\photons{photons}
\DeclareSIUnit\atoms{atoms}
\begin{document}
\title{Measurement-enhanced determination of BEC phase transitions}
\author{Mark G. Bason}
\thanks{M.~G.~Bason and R.~Heck contributed equally to this work}
\affiliation{Department of Physics and Astronomy, Aarhus University, DK-8000 Aarhus C, Denmark}
\affiliation{Present address: Department of Physics and Astronomy, University of Sussex, Falmer, Brighton BN1 9QH, UK}
\author{Robert Heck}
\thanks{M.~G.~Bason and R.~Heck contributed equally to this work}
\author{Mario Napolitano}
\affiliation{Department of Physics and Astronomy, Aarhus University, DK-8000 Aarhus C, Denmark}
\author{Ott\'{o} El\'{i}asson}
\affiliation{Department of Physics and Astronomy, Aarhus University, DK-8000 Aarhus C, Denmark}
\author{Romain M\"{u}ller}
\affiliation{Department of Physics and Astronomy, Aarhus University, DK-8000 Aarhus C, Denmark}
\author{Aske Thorsen}
\affiliation{Department of Physics and Astronomy, Aarhus University, DK-8000 Aarhus C, Denmark}
\author{Wen-Zhuo Zhang}
\affiliation{Department of Physics and Astronomy, Aarhus University, DK-8000 Aarhus C, Denmark}
\affiliation{Present address: Innovation Center of Quantum Information and Quantum Physics, Chinese Academy of Sciences, University of Science and Technology, Shanghai, China.}
\author{Jan Arlt}
\affiliation{Department of Physics and Astronomy, Aarhus University, DK-8000 Aarhus C, Denmark}
\author{Jacob F. Sherson}\email[Electronic address: ]{sherson@phys.au.dk}
\affiliation{Department of Physics and Astronomy, Aarhus University, DK-8000 Aarhus C, Denmark}

\date{\today}

\begin{abstract}
We demonstrate how dispersive atom number measurements during evaporative cooling can be used for enhanced determination of the parameter dependence of the transition to a Bose-Einstein condensate (BEC). In this way shot-to-shot fluctuations in initial conditions are detected and the information extracted per experimental realization is increased. 
We furthermore calibrate in-situ images from dispersive probing of a BEC with corresponding absorption images in time-of-flight. This allows for the determination of the transition point in a single experimental realization by applying multiple dispersive measurements. Finally, we explore the continuous probing of several consecutive phase transition crossings using the periodic addition of a focused ``dimple'' potential.
\end{abstract}
\pagestyle{plain}
\maketitle

Quantum effects have become increasingly important in the development of modern technologies and have led to the need for accurate quantum simulation~\cite{Georgescu2014,Bloch2012a}. Various approaches, including cold atoms~\cite{Dalibard2011,Simon2011,Struck2011} and ions~\cite{Britton2012}, superconducting circuits~\cite{Houck2012}, and photons~\cite{Aspuru-Guzik2012} are used to perform such simulations. In particular, many quantum simulators aim to pin down the boundaries between discrete phases of many-body systems, such as Ising spin transition points~\cite{Simon2011,Britton2012} or magnetic phases~\cite{Struck2011}, with the highest possible accuracy.  Current simulations focus on providing tight experimental bounds for benchmarking theoretical models, such as finite temperature bosonic superfluids in optical lattices~\cite{Trotzky2010a}.

In this article we investigate the potential for using dispersive probing of ultracold atom clouds in two distinct toy-model settings to enhance future quantum simulations. First, we demonstrate that benchmark probes of thermal clouds during forced evaporative cooling can be used to enhance the accuracy in determining the {\emph{critical point}}, at which the transition to a Bose-Einstein condensate (BEC) occurs. Second, we investigate multiple, in-situ probing of BECs during evaporation as a means of single-shot mapping of a phase transition.   

Since the pioneering work on dispersive probing~\cite{Andrews1996,Bradley1997}, it has been employed to investigate the relative repulsion of BECs and thermal clouds~\cite{Meppelink2010} and to monitor the formation of a BEC~\cite{Miesner1998,Wigley2016} as well as the appearance and dynamics of solitons~\cite{Lamporesi2013,Nguyen2014}. In-sequence benchmark measurements of cold atomic clouds have recently been applied as a tool to moderately reduce classical fluctuations in atom number~\cite{Sawyer2012}. Using active feedback, atom number stabilization of cold clouds to the $10^{-3}$~level was demonstrated in Ref.~\cite{Gajdacz2016}. However, to date no experiments have explored benchmark-measurement enhanced evaluation of the BEC atom number. Efficient suppression of classical fluctuations could enable the settlement of long-standing theoretical disputes concerning the nature of fundamental fluctuations arising at the BEC phase transition~\cite{Kocharovsky2006}. 

In recent years, theoretical studies on dispersive light-matter interaction based Quantum Non-Demolition (QND) measurements have demonstrated these as a useful tool for the detection and control of quantum many-body systems. Quantum noise-limited probing and feedback may enable the line-width reduction of atom lasers~\cite{Szigeti2010}, the stabilization of BECs against spontaneous emission~\cite{Hush2013}, and to squeeze and entangle Bogoliubov excitations~\cite{Wade2015}. If extended to optical lattices, it may allow for QND detection of quantum phase transitions~\cite{Eckert2007,Rogers2014} as well as Schr\"{o}dinger cat state generation~\cite{Pedersen2014,Mazzucchi2016} and arbitrary quantum state engineering using tunneling~\cite{Pedersen2014} and pinned~\cite{Hauke2013} atoms. 
To realize its full potential, however, careful studies of the fundamental information-disturbance trade-off~\cite{Lye2003,Gajdacz2013,Ilo-Okeke2014,Bons2016,Wigley2016} associated with dispersive imaging are required.

This article is structured as follows: In Sec.~\ref{s:benchmark_probe}, we show that minimally destructive measurements of the atom number early in the experimental sequence can be used to predict the BEC atom number. Based on the results of these dispersive measurements of the intermediate phase-space density (PSD) we group the BEC data and demonstrate a clear shift in the BEC transition versus initial conditions. We use this additional information for a high-resolution investigation of the transition point in our toy-model. The purpose of this investigation is not to obtain parameters of this well-studied transition at high precision but rather to demonstrate a novel methodology to increase the available information per experimental run. This methodology may also find applications for fluctuating magnetic fields in the determination of magnetic order phase transitions in Ising spin chains~\cite{Simon2011}, impurity doping concentrations for the quantum simulation of the doped Hubbard Hamiltonian~\cite{Mazurenko2017,Keimer2017}, or the chemical potential in the realization of both Bose- and Fermi-Hubbard models. Recently, post-selection of data conditioned on in-sequence insitu measurements of magnetic field values was used to enhance Rabi oscillation contrast~\cite{Krinner2017}.

In Sec.~\ref{s:dispersive_probing} of this article, we perform a detailed comparison of images obtained by dispersively probing BECs in-situ and equivalent absorption images after a time-of-flight. This allows for the single-shot determination of the phase transition using multiple dispersive probe pulses as well as the quantification of the deterministic shift of the critical point due to probe induced heating. Finally, we take first steps to dispersively probe  several consecutive  phase transitions using the periodic addition of a focused \emph{dimple} potential in a single experimental realization.

\section{Faraday imaging as dispersive benchmark probe}
\label{s:benchmark_probe}
 \begin{figure}[!ht]
		\includegraphics[width=\columnwidth]{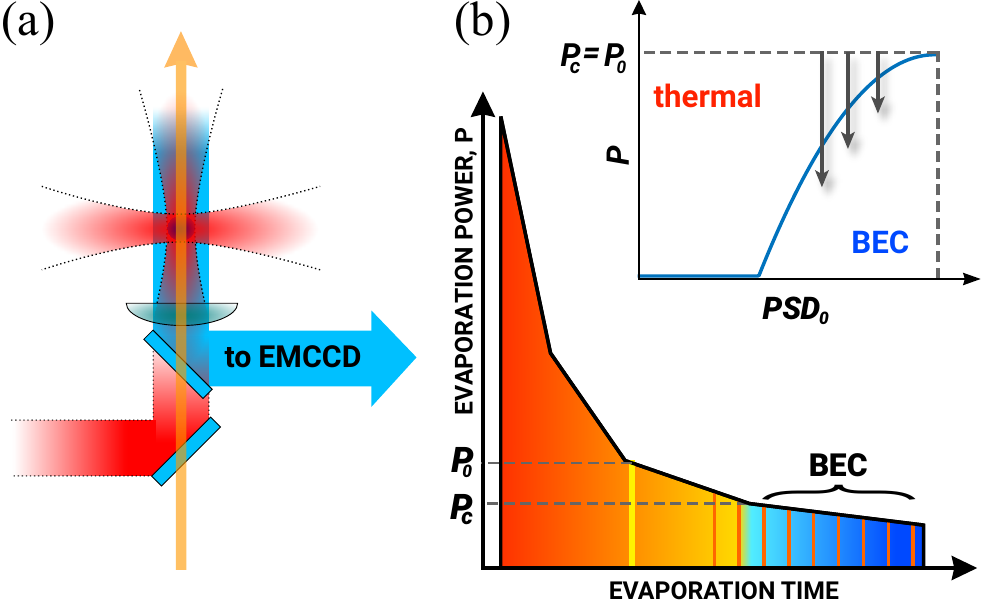}
	\caption{(Color online) (a)~Experimental setup showing the crossed dipole trap (red) at a wavelength of 1064 nm and the imaging lens. Detuned light (blue) is used to dispersively probe the BEC. A dimple potential (orange arrow) at a wavelength of \SI{912}{\nano\meter} allows for repeated phase transition crossings. The probing light and the trapping light are separated by dichroic mirrors. A polarizing beam splitter (not depicted) filters the rotated component of the probing light. (b)~Evaporation scheme showing the evaporation power as a function of time. Illustrated are the dispersive benchmark pulse at evaporation power $P_o$ (yellow line) and the probing around the phase transition (red lines) realized either with destructive absorption imaging or with multiple dispersive probe pulses. The inset schematically shows the toy-model {\textit{critical curve}} describing the  power, $P_c$,  at which the transition to a BEC occurs as a function of the phase space density, \PSDo, measured dispersively at  $P_o$. The arrows illustrate the paths of individual experimental realizations for different \PSDo.}
	\label{fig:overview}
\end{figure}

Our experimental setup is shown in Fig.~\ref{fig:overview}(a). Atoms are prepared in the $\ket{\textrm{F} = 2, \textrm{m}_\textrm{F}=2}$ state of $^{87}$Rb and loaded into a crossed dipole trap consisting of two orthogonal, focused \SI{1064}{\nano\meter} laser beams with waists of \SI{70}{\micro\meter}. Forced evaporative cooling is performed by lowering the power in each beam in a sequence of linear ramps from an initial evaporation power of \SI{5.5}{\watt} to final values between $900$ and \SI{600}{\milli\watt}, where partially condensed clouds with condensed fractions $N_\mathrm{0}/N\!\approx\,$\numrange[range-phrase = --]{0}{0.6} and total atom numbers $N\!\approx\,$\numrange[range-phrase = --]{2.1}{0.8}$\cdot10^{6}$ are produced. 

Non-destructive, dispersive measurements of the cloud are recorded using Dark Field Faraday Imaging (DFFI) \cite{Gajdacz2013}. In this technique the Faraday effect leads to a rotation of the incoming linear light polarization proportional to the column density of an atomic cloud. To implement DFFI, probing  light blue detuned by $\Delta$ = \SI{1.5}{\giga\hertz} from the $\ket{\textrm{F} = 2}\rightarrow\ket{\textrm{F'} = 3}$ transition propagates along the same axis as one of the dipole trap beams. Intensities of \SI{10}{\milli\watt\per\centi\meter\squared} are used, corresponding to around \SI{400}{\photons\per\micro\meter\squared\per\micro\second}. A small bias magnetic field of around \SI{1}{\gauss}  along the light propagation direction preserves the atomic sample's spin polarization. By inserting a polarizing beam splitter  cube with a suppression factor of \num{3e-4} in the imaging path, only the rotated component of the light is imaged onto an Electron Multiplying Charge Coupled Device (EMCCD) camera  with an effective spatial resolution of \SIrange[range-phrase = --, range-units=single]{3}{4}{\micro\meter}. 
In each realization of the experiment, a DFFI pulse of \SI{20}{\micro\second} duration probes the thermal cloud at a trapping power $P_o=\SI{1.1}{\watt}$~\footnote{Different values of $P_o$ have been investigated. The chosen one allows for the most precise determination of \PSDo, while the influence of the light pulse on the atom cloud is minimal.} (see Fig.~\ref{fig:overview}(b)) and thus realizes a dispersive benchmark (DB) measurement. The total atom number and temperature is extracted from this in-situ image: typical values are around \SI{3e6}{\atoms} at \SI{1}{\micro\kelvin} giving a reference phase space density \PSDo$\,\approx\,0.25$. The influence of this pulse is at the 1\% level in terms of heating and even lower in terms of atom losses. Although higher precision probing can be realized by increasing the probing intensity~\cite{Gajdacz2016}, this minimally destructive probing was chosen to allow for the exploration of the maximal possible range of condensate fractions. 
 
Thus, a reference PSD is known for each run of the experiment before the phase transition occurs. Assuming that the subsequent evaporation efficiency is deterministic, the critical dipole trap power, $P_c$,  at which the phase transition is crossed can be predicted. A high \PSDo\ will lead to efficient evaporation and thereby a phase transition at high optical power. If \PSDo\ is reduced, $P_c$ reduces until a BEC is no longer obtained below a certain  \PSDo. The dependence of $P_c$ on \PSDo\ defines a {\textit{critical curve}} illustrated in the inset of Fig.~\ref{fig:overview}(b). In the following we use this example to demonstrate the benefits of dispersive probing for the determination of phase diagrams in the presence of environmentally induced fluctuations of e.g.\ laser beam intensity.

\begin{figure}[tb]
		
		\includegraphics[width=1\columnwidth]{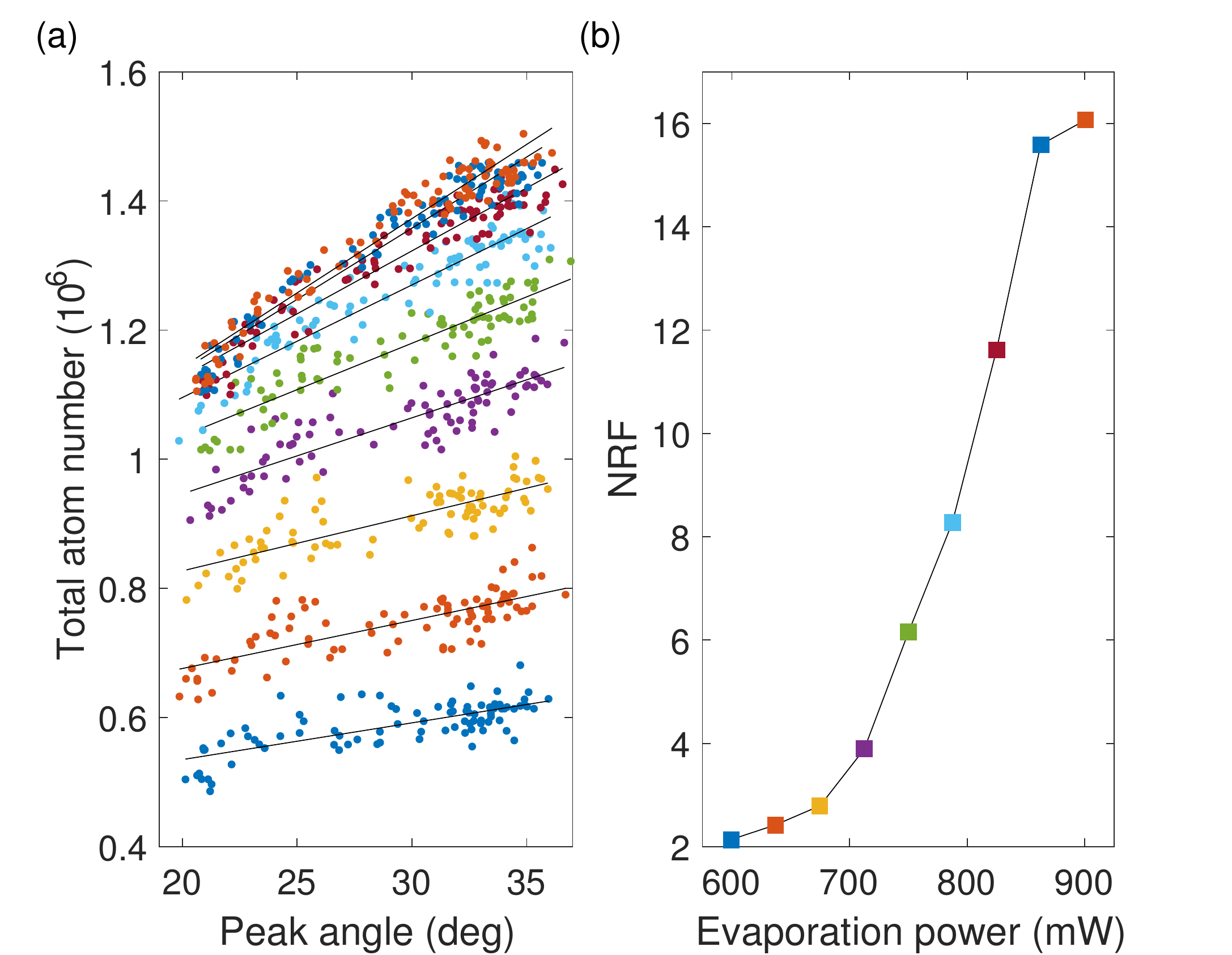}

\caption{(Color online) (a) Correlation between the peak rotation angle from dispersive benchmark probing and the total atom number from absorption imaging for various final trap depths (from top to bottom: higher to lower final evaporation power). The solid black lines are linear fits to the data. (b)~The extracted NRFs are shown for the different final evaporation powers. The solid black line serves as guide to the eye.}
\label{fig:nsum_peak-angle_correlation}
\end{figure}

We first quantify the predictive power of our DB method by analyzing fully destructive absorption images acquired at 9 different final powers during the last evaporation stage across the BEC transition between $900$ and \SI{600}{\milli\watt}. At each point, the variance of the absorption measurements, $\sigma^2_y$, is compared to the reduced variance, $\sigma^{2}_{y,\mathrm{red}}$, given the knowledge of the benchmarking probe. 
The correlation of the peak rotation angle of a DB pulse and the total atom number inferred from absorption imaging is shown in Fig.~\ref{fig:nsum_peak-angle_correlation}(a). We define the noise reduction factor (NRF) for a general measured quantity $y(x)$ as the ratio between the original and reduced variances:
\begin{align}
    \textrm{NRF} &= \frac{\sigma^2_y}{\sigma^{2}_{y,\mathrm{red}}},\\
    \sigma^{2}_{y,\mathrm{red}} &= \frac{1}{N-1}\sum_{i=1}^N \left[ y_i-\bar{y}(x_i) \right] ^2,
\end{align}
where $\bar{y}(x)$ is determined by simple linear regression. Hence, stronger correlations result in higher values of NRF.

The associated NRFs are shown in Fig.~\ref{fig:nsum_peak-angle_correlation}(b). Slightly before the transition point, we obtain a reduction in variance by a factor of 16 indicating a strong correlation between the two variables. After the transition point, however, the reduction decreases significantly and ends at roughly 2 for the data points at the lowest final evaporation power, \SI{600}{\milli\watt}. This drastic drop in predictability is attributed to a combination of small classical fluctuations, such as trap bottom instabilities, and quantum noise arising due to the stochastic nature of the condensation formation~\cite{Kocharovsky2015}. The drop demonstrates that \emph{single} benchmark measurements cannot give detailed information on the stochastic dynamics in the region around the BEC transition. Nonetheless, the simplicity of this technique, makes it a powerful tool and it can still be used to enhance the precision in determination of the critical point. 

  \begin{figure}[t]
 		\includegraphics[width=1\columnwidth]{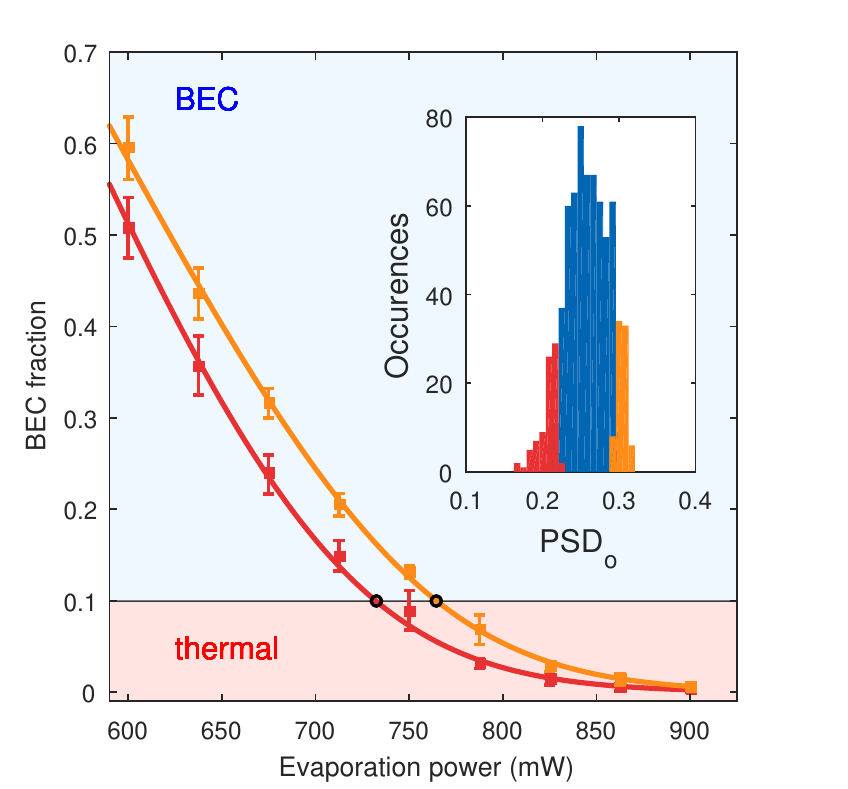}
 	\caption{(Color online) BEC growth curves from absorption imaging data for two different data groups of \PSDo. The solid lines are fits with a heuristically growth motivated function (Eq.~\ref{eq:BEC_growth_function}), where the black circles mark the critical power, $P_c$ (see text for details). The error bars represent one standard deviation. The inset shows the histogram of \PSDo\ measured at \SI{1.1}{\watt}. The spread in \PSDo\ originates from atom number and temperature fluctuations in the experiment. The red and yellow shaded areas depict the two selected data groups for the respective growth curves.}
 	\label{fig:tofdata}
 \end{figure}

The predictive power of our DB method is utilized by binning  different experimental runs in which \PSDo\ is approximately equal. To limit the systematic error due to inaccurate analysis of absorption images at small condensate fraction, we experimentally define the critical power, $P_c$, as the point at which the BEC fraction reaches \SI{10}{\percent} for the remainder of this article. $P_c$ in each bin is then determined by fitting the condensate fraction with the heuristically motivated function
\begin{align}
N_\mathrm{0}/N= \alpha (P-\beta)/[1-\exp(\gamma (P-\beta))],
\label{eq:BEC_growth_function}
\end{align}
where $P$ denotes the final evaporation power and $\alpha$, $\beta$ and $\gamma$ are fitting parameters.  For illustration, we first apply this procedure to two groups of data points (see inset of Fig.~\ref{fig:tofdata}). As expected, a clear systematic shift between the two groups is observed. 

  \begin{figure}[!ht]
 		\includegraphics[width=1\columnwidth]{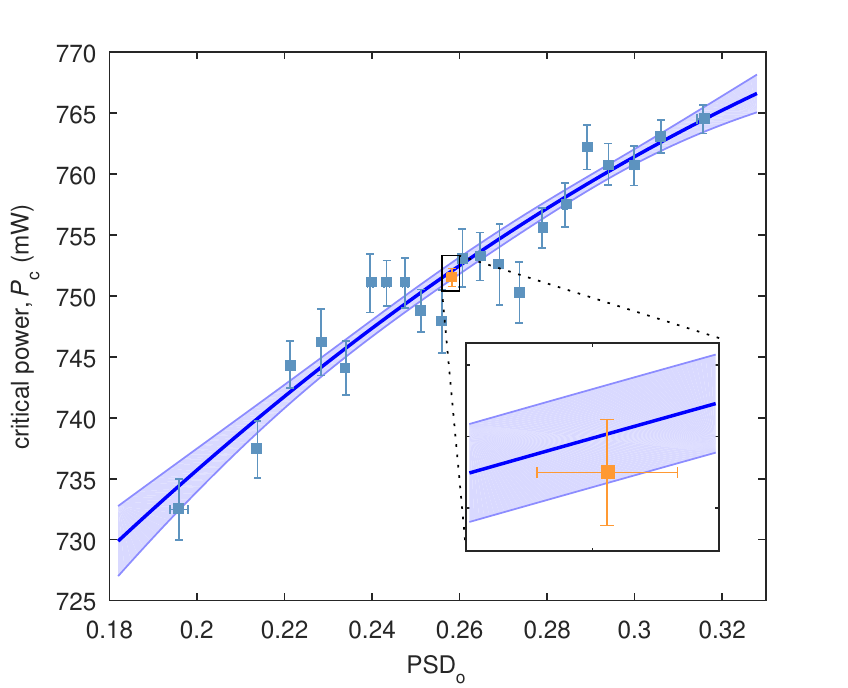}
 	\caption{(Color online) Experimental determination of the critical. Each blue point is obtained from 4 individual evaporation series.  The horizontal and vertical error bars correspond to the standard error of \PSDo\ and $P_c$ in each bin, respectively. The solid blue curve is a quadratic fit to the data, where the shaded area indicates the $1\sigma$ confidence bounds. The yellow point is the result of evaluating the whole data set without binning. The inset (\PSDo~$=\,$\numrange[range-phrase = --]{0.256}{0.26},\ $P_c\!=\,$\SIrange[range-phrase = --]{750.4}{753.3}{\milli\watt}) highlights the area around the point representing the whole dataset.}
 	\label{fig:PSDvsPc}
 \end{figure}

In a more systematic analysis we examine how much information about the shape of the critical curve can be extracted. The data is divided into 21 bins corresponding to 21 different mean values of \PSDo. For each bin the following analysis is performed: Full data sets of BEC fractions as a function of evaporation power are selected randomly and fit with the growth function (Eq.~\ref{eq:BEC_growth_function}). This is repeated until all available data of the bin under investigation are used. Out of the obtained set of fit results, the mean value and the standard error of $P_c$ are extracted. The result, as a function of \PSDo, is displayed in Fig.~\ref{fig:PSDvsPc} (blue data points). 
The DB method makes use of the classical fluctuations of \PSDo\ to explore an extended regime of the critical curve in a single measurement series. Conventionally, such critical curves are investigated by manually choosing different settings of the control variable for each measurement series. The fact that the control variable is normally only known up to the classical fluctuations means that significantly more data has to be acquired in order to resolve the critical curve with a quality comparable to the one achieved using the DB method. 

In addition, the conventional data averaging method usually involves implicit assumptions of the shape of the curve to be investigated. 
In general, if there is no \emph{a priori} knowledge of the shape of the critical curve or the distribution of the fluctuations of the control variable, a conservative estimate would have to employ the \emph{standard deviation} for extracting uncertainties. 
Conventionally, the fluctuations are assumed to explore a linear part of the critical curve and all the data is represented by the standard error as is done for the orange point in Fig.~\ref{fig:PSDvsPc}. The figure inset demonstrates that the precision of our method is comparable while simultaneously yielding information about the shape of the curve. Moreover, if fluctuations were significant enough to explore a slightly quadratic part of the curve, the combined standard error analysis would  lead to inaccurate and misleading error bounds (see also appendix). We emphasize the relative simplicity of the technique, which can easily be implemented in existing setups as a post-selection tool or to study the stochastisticity of quantum phase transition dynamics in general. 

\section{Dispersive probing of the BEC transition}
\label{s:dispersive_probing}

Having established dispersive probing as a useful benchmark tool, we now investigate the use of individual and multiple DFFI pulses to interrogate the BEC transition directly. So far, dispersive measurements have been used to realize squeezing of cold but uncondensed atomic clouds using photo detectors~\cite{Appel2009,Takano2009,Koschorreck2010,Schleier-Smith2010}. 
However, these approaches do not provide any spatial information about the atomic cloud. Spatially resolved, quantum-noise limited probing would be essential to create squeezed and entangled states of Boguliobov excitations within a BEC~\cite{Wade2015}.

Similarly, the probing of the BEC transition requires spatial resolution to be able to fit density profiles. The ultimate goal is accurate probing of the transition at the quantum limit. This may enable access to quantum fluctuations in two cases: First, the characterization of the stochastic onset time of the spontaneous symmetry breaking process at condensation similar to the stochastic  initial appearance of pair creation in a spinor condensate at $m_F=0$~\cite{Klempt2010}. Second, the observation of condensate fluctuations for a gas with a defined total atom number~\cite{Kocharovsky2006}. Classical fluctuations have so far prevented the observation of such quantum fluctuations~\cite{Miesner1998}.

Although initial efforts to understand the spatial two-component structure of the partially condensed clouds have been undertaken~\cite{Meppelink2010}, diffraction effects, the full numerical multi-mode modelling of interaction effects, finite photon number effects in the detection, and classical heating effects of the probe itself are still not fully understood. In the following we investigate the latter two. 

This section is structured as follows: in Sec.~\ref{section:fitting}, we describe the fitting procedure for extracting the BEC atom number from in-situ DFFI images. Low photon flux leads to a systematic shift in the obtained BEC fractions which we confirm using simulated density profiles presented in Sec.~\ref{section:accuracy}. Sec.~\ref{section:transition_shift} demonstrates how classical heating induced by the probe itself influences the critical point. Finally, we show progress towards the first attempt of monitoring multiple, conservative crossings of the BEC transition probed in a single experimental realization using DFFI pulses. 

\subsection{Density model and fitting technique}
\label{section:fitting}
In contrast to well understood time-of-flight absorption imaging where atomic densities are low and the mean-field interaction between atoms is negligible, in the case of in-situ images the influence of interaction on the cloud's density profile has to be taken into account. The exact shape of in-trap clouds is a matter of current investigation and model dependent~\cite{Meppelink2010}.

Here, we employ the \emph{semi-ideal} model~\cite{Naraschewski1998,Minguzzi1997,Gerbier2004,Meppelink2010} which, instead of taking the complete influence of mean-field interactions on the condensed and thermal part into account in a self-consistent way, only includes the effect of the condensed part on the thermal part and leads to a modified effective trapping potential for the thermal part
\begin{equation}
 V_\mathrm{eff}(\mathbf{r}) =  V_\mathrm{ext}(\mathbf{r}) + 2gn_0(\mathbf{r}),
\label{eq:effective_potential_popov}
\end{equation}
with $V_\mathrm{ext}$ being the external trapping potential, $n_0$ the density distribution of the condensed part and $g = \frac{4\pi \hbar^2 a}{m}$ the coupling constant. Here, $a$ is the $s$-wave scattering length and $m$ the atomic mass.
 
In Thomas-Fermi approximation, the condensed and thermal cloud's density distributions are then given by 
\begin{align}
n_0(\mathbf{r}) &= \frac{1}{g}\left(\mu - V_\mathrm{ext}(\mathbf{r})\right)\quad\text{and}
\label{eq:condensate_density_semi-ideal}\\
n_\mathrm{th}(\mathbf{r}) &= \frac{1}{\lambda_\mathrm{T}^3}g_{3/2}\bigl(e^{\beta(\mu- V_\mathrm{eff}(\mathbf{r}))}\bigr),
\label{eq:thermal_density_semi-ideal}
\end{align}
where $\lambda_\mathrm{T}$ is the thermal de Broglie wavelength and the Bose function $g_{3/2}(z)$ generally defined as $g_p(z) = \sum_{l=1}^{\infty}z^l/l^p$. Usually, the chemical potential $\mu$ and the number of condensed atoms $N_0$ have to be determined self-consistently. We circumvent this by applying a further simplification and expanding the condensed fraction into a series depending on the reduced chemical potential $\bar{\mu} = \frac{\mu}{k_BT}$ \cite{Naraschewski1998}. Truncation after the first non-trivial order and solving for the condensate fraction yields
\begin{equation}
\frac{N_0}{N} = 1-\left(\frac{T}{T_c}\right)^3-\eta\frac{\zeta(2)}{\zeta(3)}\left(\frac{T}{T_c}\right)^2\left[1-\left(\frac{T}{T_c}\right)^3\right]^{2/5},
\label{eq:condensate_fraction}
\end{equation}
where $\eta$ is a dimensionless scaling parameter given by $\eta= \frac{\mu_{T=0}}{k_BT}$ and $T_c$  the critical temperature for the onset of condensation for a non-interacting, ideal gas $k_BT_c = \frac{\hbar\bar{\omega} N^{1/3}}{\zeta(3)^{1/3}}$, with the Riemann zeta function $\zeta$. 

Thus, for a cloud with atom number $N$ at temperature $T$, one can calculate the expected condensate fraction and the density distributions for thermal and condensed parts.

  \begin{figure}[tb]
 		\includegraphics[width=1\columnwidth]{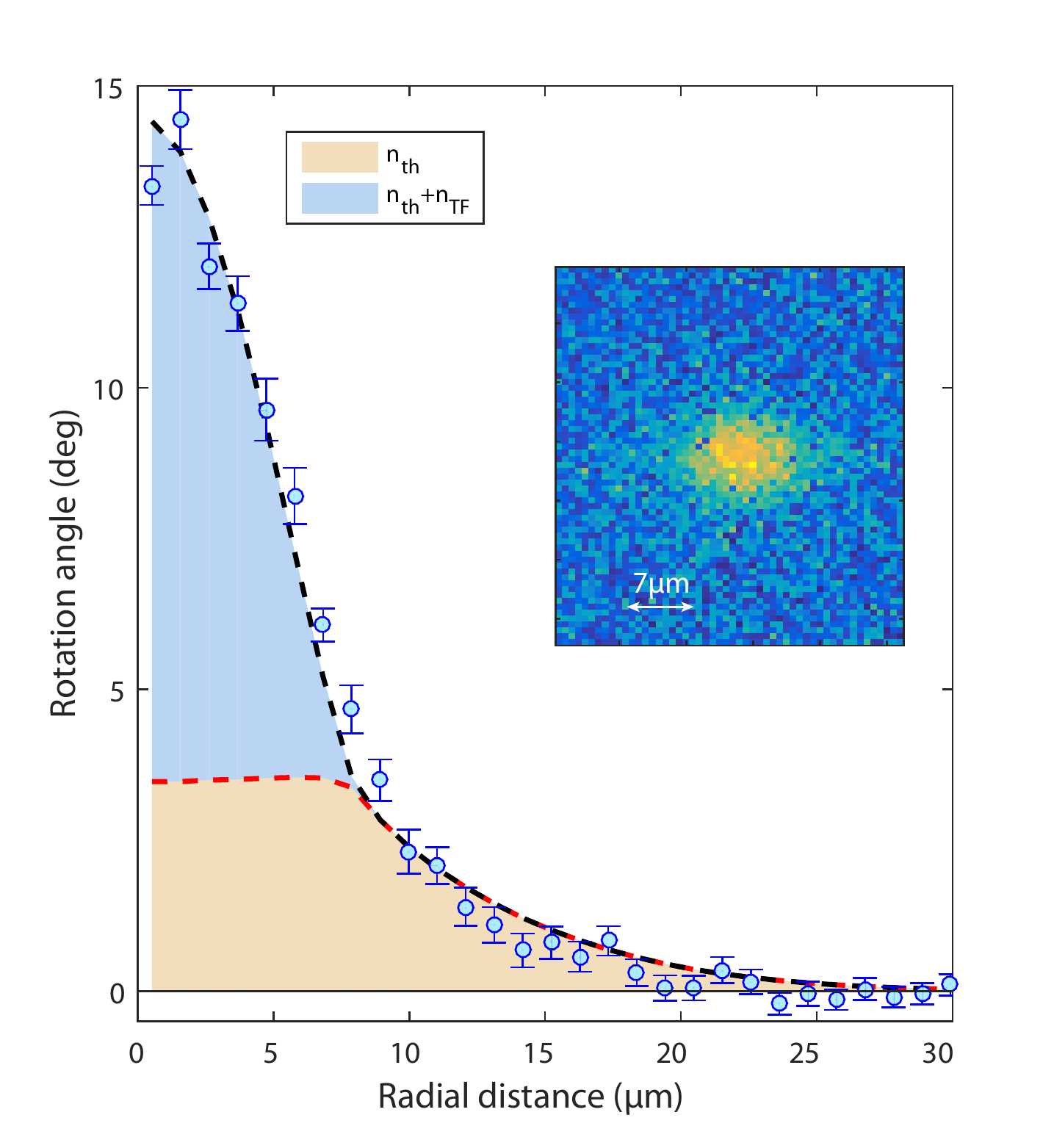}
 	\caption{(Color online) Azimuthal averaging and fitting with the semi-ideal model. The raw image (shown in the inset) is azimuthally averaged around the center of the cloud (blue data points, the error bars depict the standard deviation obtained in the averaging process). The averaged data is compared to results of the semi-ideal model and the best agreement is found. The red dashed line shows the thermal atom distribution of the best fitting realization, the blue dashed line indicates the corresponding sum of thermal and condensed atom distribution.}
 	\label{fig:fitting_collage}
 \end{figure}

To extract information about atom number, temperature and BEC fraction from the dispersively probed in-situ cloud, each image is first converted from EMCCD counts into rotation angles~\cite{Gajdacz2013}. The resulting image is azimuthally averaged around the point with the highest rotation angle (see Fig.~\ref{fig:fitting_collage}). This radial profile is then compared to multiple profiles obtained for realizations of the semi-ideal model with atom clouds of different $N$ and $T$ and a given trap geometry. The profile with the minimal sum of squared, normalized residuals 
$\chi^2$ is found and thus the best fitting pair of $N$ and $T$ is obtained. 
The obtained peak densities are within 20\% of those extracted from absorption images. 

A problem arises for clouds which are just above the condensation threshold, since there is a range of temperatures at which a BEC is predicted for a non-interacting gas, but not according to the semi-ideal model. In order to close this gap, we linearly interpolate the density distributions between the cases of a small interacting BEC and a non-interacting thermal cloud. 

The 1$\sigma$ confidence bounds on $N$ and $T$ are found by identifying the points at which $\chi^2$ takes the value $\chi^2_{\mathrm{min}}+1$~\cite{Hughes2010}. Using this method, the statistical errors for the condensate fraction based on a single fit are on the level of 1\%. 

\subsection{Accuracy of DFFI images close to the BEC threshold}
\label{section:accuracy}
\begin{figure}[tb]
 		\includegraphics[width=1\columnwidth]{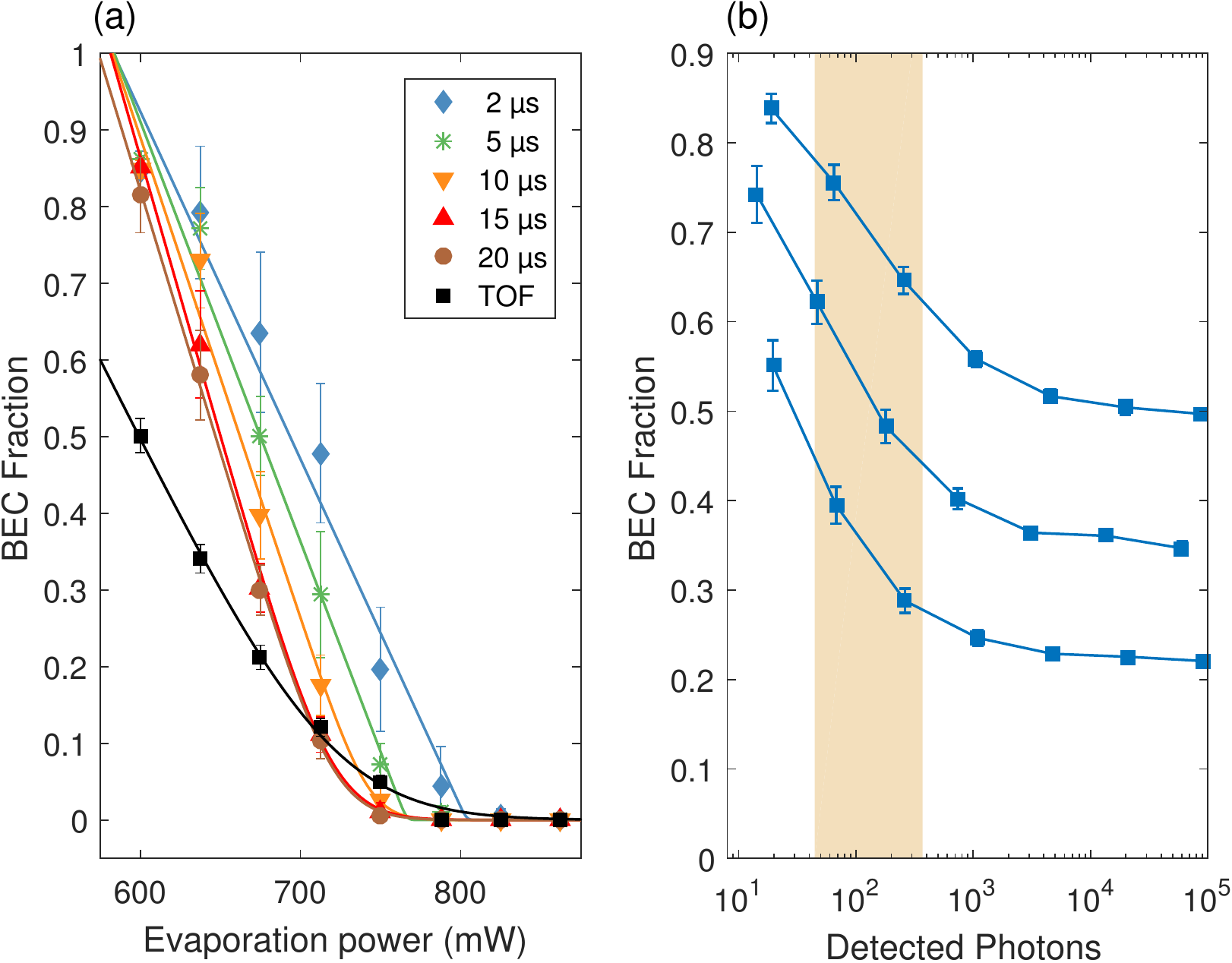}
 \caption{(Color online) (a) Single Faraday probing of the BEC transition for different imaging pulse durations (colored symbols) compared to results of conventional absorption imaging (black squares). The error bars depict one standard deviation, whereas the coloured solid lines are fits with the growth function of Eq.~\ref{eq:BEC_growth_function}. The systematic overestimation of BEC fractions obtained in DFFI is visible.  (b)~Simulation of the influence of shot noise on the fit results in DFFI as a function of mean number of detected photons. For the modelling, realistic pairs of total atom number $N$ and temperature $T$ were chosen corresponding to three different BEC fractions (from top to bottom $\frac{N_0}{N} = 0.50$, $0.35$ and $0.23$). The lines are guides to the eyes. A clear systematic shift of the BEC fraction is observed for photon numbers below $10^3$. The shaded area indicates the photon numbers used in our experiments.}
 \label{fig:Faraday_singleprobe}
 \end{figure}

To examine the effect of Faraday probing, a single dispersive pulse of varying duration is applied for a range of potential depths across the BEC transition. BEC fractions from the fitted profiles are then compared to those measured for the same potential with absorption imaging. This analysis, shown in Fig.~\ref{fig:Faraday_singleprobe}(a), shows a systematic shift towards higher BEC fractions in the in-situ images.
 
To understand the origin of this trend the effect of finite photon numbers on the determination of BEC fractions is modelled. For a given photon flux and cloud profile sample images, which include the Poissonian statistics of the imaging light, are generated and the fitting procedure described previously is applied. The mean and standard deviation in BEC fraction are then extracted from the series of images and repeated for different photon fluxes. The results, shown in Fig.~\ref{fig:Faraday_singleprobe}(b), demonstrate a clear, systematic shift consistent with observations. This effect persists for all the three simulated condensate fractions. The range of photon fluxes used in the experiment (shaded area) indicate that agreement between in-situ and absorption imaging cannot be expected for higher condensate fractions. This effect is attributed to low signal-to-noise in the thermal wings of the cloud such that parts are below the detection limit. Therefore, the clouds appear to be purer in terms of condensate fraction.

\subsection{Probing dependent shift of the critical power $P_c$}
\label{section:transition_shift}
Building upon the understanding of the individual in-situ images, we repeatedly probe the sample in a single evaporative run. Here we address an important question: the modification of the phase transition due to the heating caused by the probing. This is important for future studies due to the intrinsic trade-off between destructivity and signal to noise in dispersive probing.

\begin{figure*}[!ht]
		\includegraphics[width=\textwidth]{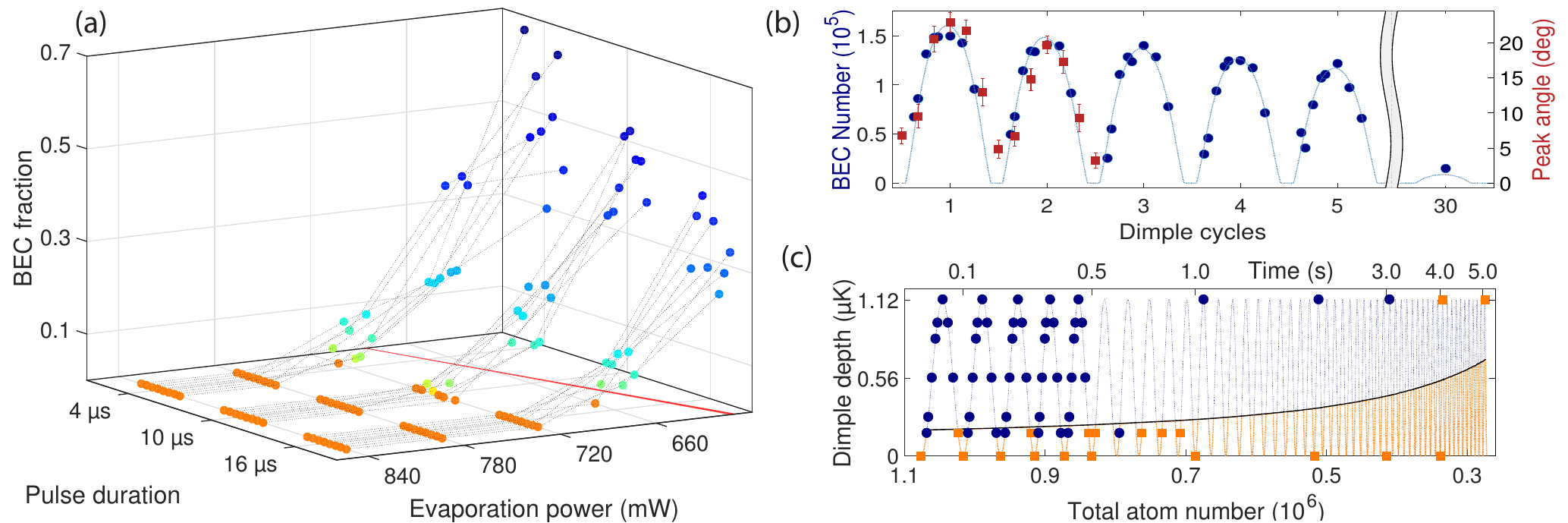}
	\caption{(Color online)  (a) Single-shot probing of the BEC formation: For each of the three pulse durations 8 repetitions are shown. The red line, derived from absorption images, indicates the critical power $P_c$ (thickness corresponds to $1\sigma$ confidence bounds). (b) BEC atoms measured via absorption imaging (blue circles) as the dimple trap depth was cycled. Red squares show the peak rotation angle from multiple dispersive images of 2 $\mu$s duration during the first two cycles; error bars correspond to one standard deviation over several repetitions. The solid line is a guide to the eye. (c) Phase-diagram of the atoms in the dimple potential. Above the black line a BEC should be observed (blue part of the sinusoidally cycling curve), while below one expects only thermal atoms (orange part). The blue dots indicate absorption  measurements containing a BEC, while the orange squares indicate purely thermal clouds.}
	\label{fig:dimple}
\end{figure*}

During the final stages of the evaporation process up to 6 light pulses are taken at \SI{220}{\milli\second} intervals for various pulse durations. After the final pulse in the sequence an absorption image is taken and the condensed fraction extracted. The results of this in-situ probing of the condensation process using three different probe durations are shown in Fig.~\ref{fig:dimple}(a)~\footnote{The results of the dispersive probing presented in the following were not corrected for the observed systematic shift due to low photon flux.}. A clear shift of the BEC transition point towards lower dipole powers and thus shallower traps is observed for increasing pulse durations due to heating. This shift is confirmed independently by the absorption images taken at the end of the sequence. As in Sec.~\ref{s:benchmark_probe}, we use the same heuristically motivated function to extract the critical power $P_c$ where 10\% condensate fraction is reached. A linear fit to $P_c$ as a function of the pulse duration is depicted as the red curve in Fig.~\ref{fig:dimple}(a). The observed heating rates due to the dispersive probing are somewhat higher than simple theoretical estimates suggest, which we attribute both to classical variations in probing conditions and to multiple scattering events due to the high spatial density. Our results reinforce the need to consider heating effects in future studies of the quantum effects of dispersive probing~\cite{Bons2016,Wigley2016}.
\subsection{Multiple crossing of the phase transition}
\label{section:multiple_crossing}
Finally, we take first steps towards the single run mapping of an entire phase diagram by combining repeated conservative crossings of a phase transition with continuous dispersive probing. In particular, periodic addition of a focused `dimple' potential~\cite{Pinkse1997} allows for the repeated crossing between thermal cloud and BEC. Studies to date have only employed this effect in combination with individual, destructive absorption images: BEC production was characterized \cite{Comparat2006a,Garrett2011,Dutta2015} and in one pioneering experiment multiple crossings explored~\cite{Stamper-Kurn1998}. 

Our experiment is conducted as outlined in Sec.~\ref{s:benchmark_probe}, but the evaporation is stopped before condensation is reached. At this point, the cloud serves as a reservoir in which a \SI{912}{\nano\meter} wavelength laser beam focused to a waist of \SI{7}{\micro\meter} produces an additional potential. This potential is sinusoidally cycled from a depth of $0$ to \SI{1.12}{\micro\kelvin} at a rate of \SI{10}{\hertz} to repeatedly produce a BEC in the dimple potential. We first characterize the cloud evolution with absorption imaging. Maximum BEC atom numbers of around 1.5$\cdot10^5$~atoms, 10\% of the total atom number, are initially measured while after 30~cycles condensed clouds of 2$\cdot10^4$~atoms are still distinguishable from the thermal reservoir, see the blue dots in Fig.~\ref{fig:dimple}(b). 

The condensation dynamics in the dimple involves three components of the sample: the thermal cloud in the reservoir, and the BEC and thermal components in the dimple. Due to our limited imaging resolution we cannot distinguish between the latter two. However, treating the dimple as a small perturbation, we can use the atom numbers within the dimple volume, the reservoir temperatures and the time-varying trapping frequencies of the cycling dimple potential to estimate critical dimple depths at which the BEC transition is observed.

Absorption imaging data is used to extract the evolution of the reservoir temperature and the total atom number over the 30 dimple cycles. In-situ dispersive Faraday probing, shows that up to \SI{50}{\percent} of the atoms concentrate in the volume defined by the dimple, when it is at its deepest value. 
This allows us to calculate the critical depth for each combination of atom number and reservoir temperature. 
When interactions are included, Eq.~\ref{eq:condensate_fraction} 
can be used to estimate a critical dimple depth corresponding to the threshold at which the BEC transition is crossed. Thus, a BEC phase diagram for dimple depth as a function of the total atom number can be generated, see Fig.~\ref{fig:dimple}(c). The data from the dimple cycle measurements can be sorted into two groups on this phase diagram, according to the presence or lack of a BEC. 
We find that even this simple model can predict the presence/absence of a BEC reasonably well during the first 20 cycles. The inaccuracy in the measurement of the reservoir temperature and non-adiabatic dimple loading dynamics reduce the effectiveness of the model to make predictions for the final cycles.

Subsequently, dispersive probing of multiple BEC phase transitions is performed by introducing a varying number of weak Faraday pulses within each dimple cycle at different pulse durations. In Fig.~\ref{fig:dimple}(b) red squares indicate the maximal angle of Faraday rotation obtained in each  probing pulse. The non-zero signal in the absence of the dimple is caused by the background polarization rotation of the reservoir. The data clearly demonstrate our ability to track the loading into the dimple over two realizations of the BEC production  \footnote{At present, imaging resolution does not allow to distinguish BEC and thermal parts in the dimple, however experimental reconstruction is currently underway to allow for this in the near future.}. As before, we study the destructivity of the probing by interrupting the probing at various points and determining the number of condensed atoms using absorption imaging. For a single weak probe pulse applied repeatedly at maximum dimple depth at least 10 dimple cycles can be monitored whereas only a few transitions can be monitored when seven pulses per cycle are applied. This means that both correlation measurements between, as well as high-resolution monitoring of, several BEC phase transitions in a single shot should be feasible. 

\section{Conclusion}
In summary, we have demonstrated the predictive power of non-destructively probing a cold cloud before it crosses the BEC phase transition. This provides enhanced information about the non-linear dependence of a phase transition on a control parameter. Importantly, we demonstrate that in the presence of classical shot-to-shot fluctuations using conventional methods of averaging repeated measurements the precision determining a critical point does not necessarily increase with additional repetitions unlike our method. We envision similar techniques to be applicable for any quantum many-body phase transition in which classical fluctuations play a role.
In addition, we have demonstrated repeated probing of both one and several phase transitions in a single shot. This constitutes an important first step towards the single-realization mapping of full phase diagrams as well as the investigation of stochastic effects such as increased fluctuations at the phase transition~\cite{Kocharovsky2015} and possible non-Markovian dynamics \cite{Rivas2014} for multiple crossings.

\section*{Acknowledgments}
The authors acknowledge fruitful discussions with E.~Zeuthen, J.~H.~M\"u{}ller and K.~M{\o}lmer. We thank Jens S.\ Laustsen for his feedback on the manuscript. This work was supported by the Danish Council for Independent Research,  the Lundbeck Foundation, the Villum Foundation, the European Research Council, and Marie Curie IEF. 

\label{sec:ack}

\section*{Appendix}
\label{sec:appendix}
The DB method allows for a detailed investigation of the linearity of the critical curve in the observed regime. The solid blue line in Fig.~\ref{fig:PSDvsPc} of the main text is a quadratic fit to the data and provides an upper bound for the curvature~\footnote{A $\chi^2$-test of the fit results favours the quadratic model slightly over the linear one ($\chi^2_\mathrm{red,quad}=1.20$ and $\chi^2_\mathrm{red,lin}=1.25$, respectively).}. Based on the confidence bounds of the fit ($1\sigma$, shaded areas), we obtain the standard error on the estimated $P_c$. Comparing the confidence bounds of the quadratic fit with the result of averaging all the data (orange point in the inset of Fig.~\ref{fig:PSDvsPc}), both methods give similar uncertainties at this particular point. Therefore the DB method provides an increase in the obtained information on the critical curve without a noticeable decrease in precision and is therefore especially useful for the sensitivity-enhancement of general quantum simulation experiments. 

The formal statistical agreement between the error margins for averaging over the whole data set and the confidence bounds of the fit is a reflection of the small contribution of the quadratic dependence of the critical curve within the range of \PSDo\ values.  While this component is relatively small in the case of the BEC transition, in other systems with greater non-linear dependencies on the control variables or in case of more extensive data averaging the precision would not improve with averaging of more data samples as may be conventionally assumed. 
This effect can be understood based on the following argument, which is valid for any measurement with a noisy control parameter: assuming a control parameter $x$ that is fluctuating around its mean by some amount $x'$, each realization of an experimental outcome following some function $f(x)$ will be measured at slightly different values $x=\braket{x}\pm x'$. The mean measurement result $f_{\mathrm{meas}} = \braket{f(x)}$ will simply be the convolution of the true response function with the distribution of the control variable. In the case of a normally distributed fluctuating control variable with variance $\sigma_x^2$ and assuming that the fluctuations are sufficiently small such that $f(x)$ can be approximated by its Taylor expansion to second order, one finds the introduced systematic shift,  $ f_{\mathrm{meas}} = f(x) + \frac{1}{2}f''(x)\sigma_x^2$.

A more general calculation leads to the same result even for non-Gaussian distributions of control values~\footnote{Klaus M\o{}lmer, private communication}. Since we assume a quadratic fit to our measured data the offset to the `true' result can easily be found. The whole curve is shifted by $\Delta =  -1/2 f''_\mathrm{meas}\sigma_x^2$. In the case of averaging all data this shift is $\Delta = \SI{0.25}{\milli\watt}$, while the individual binned data points have a negligible shift of $\SI{5.1}{\micro\watt}$ for the whole curve.

\end{document}